# Ultra Thin 3D Silicon Detector for Plasma Diagnostics at the ITER Tokamak

F. García[1], G. Pellegrini[2], M. Lozano[2], J.P. Balbuena[2], C. Fleta[2], C. Guardiola[2], D. Quirion[2]

[1] Helsinki Institute of Physics, University of Helsinki, Helsinki, Finland
[2] Centro Nacional de Microelectronica, Barcelona, Spain

*Abstract*–An ultra thin silicon detector called U3DTHIN[1,2] has been designed and built for neutral particle analyzers (NPA) and thermal neutron detection. The main purpose of this detector is to provide a state-of-the-art solution for detector system of NPAs for the ITER experimental reactor and to be used in combination with a Boron conversion layer for the detection of thermal neutrons. Currently the NPAs are using very thin scintillator - photomultiplier tube[3,4], and their main drawbacks are poor energy resolution, intrinsic scintillation nonlinearity, relative low count rate capability and finally poor signal-to-background discrimination power for the low energy channels. The proposed U3DTHIN detector is based on very thin sensitive substrate which will provide nearly 100% detection efficiency for ions and at the same time very low sensitivity for the neutron and gamma radiation. To achieve a very fast charge collection of the carriers generated by the ions detection a 3D electrode structure[5] has been introduced in the sensitive volume of the detector. One of the most innovative features of these detectors has been the optimal combination of the thin entrance window and the sensitive substrate thickness, to accommodate very large energy dynamic range of the detected ions. An entrance window with a thickness of tens of nanometers together with a sensitive substrate thickness varying from less than 5 μm, to detect the lowest energetic ions to 20 μm for the height ones has been selected after simulations with GEANT4. To increase the signal to background ratio the detector will operate in spectroscopy regime allowing to perform pulse-height analysis. The technology used to fabricate these 3D ultra thin detectors developed at Centro Nacional de Microelectronica in Barcelona and the first signals from an alpha source ($^{241}$Am) will presented.

## I. INTRODUCTION

THE increase in power of the plasma shots in the JET tokamak has introduced serious challenges for the operation of the Neutral Particle Analyzers (NPA) detector systems. This type of analyzers is used to perform Corpuscular Diagnostics of plasma. Such increase of the plasma burning power has increased the neutron and gamma background to the level where the detectors cannot cope with the particles rate. The detectors get saturated and are not able to detect the ions that carry the wanted information about the plasma parameters. It is expected that this problem will be even more severe in the new generation of Tokamaks, one of which will be installed in the ITER facility[6]. In order to provide a detector capable of detecting ions under such high intensity of neutrons and gamma background, a completely new detector concept using an Ultra-Thin Silicon detector with 3D electrodes is introduced.

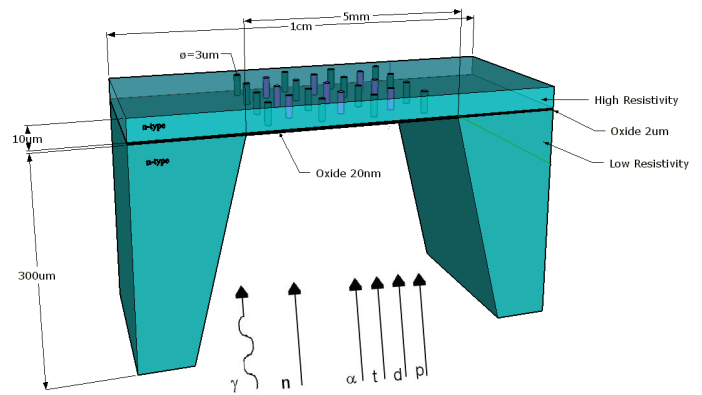

Fig. 1. Detector concept of the Ultra Thin Silicon with 3D electrodes - U3DTHIN.

This detector concept (see Fig. 1) fulfills the main requirements for the operation under high-radiation environment in terms of the count rate capability and radiation hardness. Complementary to this, the detector will have nearly 100% efficiency for detection of ions and new clusterization schemes can be explored to further improve the background rejection capability, thus increasing the signal-to-background ratio.

Simulations using GEANT4 have been carried out in order to better understand the detector performance by irradiating with background radiation (neutrons and gammas) and with ions. Complementary to this, a SENTAURUS Technology Computer Aided Design (TCAD) model has been created to study the electrical performance[7] for different geometry cells in order to get some figures of merit for the fabrication process. Finally a full fabrication run has been performed at CNM-CSIC in Barcelona.

## II. SIMULATIONS

In order to carry out a preliminary test of this detector concept, various simulation models were created. In particular, a GEANT4 model for MonteCarlo simulations of the interaction of radiation was used to obtain the energy

Manuscript received November 16, 2011. This work was supported by the Ministry of Education of Finland.

F. García is with the Helsinki Institute of Physics University of Helsinki, P.O. Box 64 FI-00014 University of Helsinki, Finland (telephone: +358-9-19151086, e-mail: Francisco.Garcia@helsinki.fi).

G. Pellegrini, M. Lozano, J.P. Balbuena, C. Fleta, C. Guardiola, D. Quirion are with the Centro Nacional de Microelectronica, Barcelona, Spain (e-mails: giulio.pellegrini@imb-cnm.csic.es, Manuel.Lozano@imb-cnm.csic.es, juanpablo.balbuena@imb-cnm.csic.es, Celeste.Fleta@imb-cnm.csic.es, consuelo.guardiola@imb-cnm.csic.es, david.quirion@imb-cnm.csic.es).

deposition by ions and background. The GEANT4 model geometry description includes all the components of the detector. The model was used to obtain accurate energy deposition values in its sensitive volume. The geometry description of the detector has the following components: a very thin entrance window of Silicon Oxide of 20 nm, a supporter silicon frame of 300 μm, the silicon-sensitive detector of 10 μm, holes of 3 μm and metallic strips on the back side made of Aluminum with a thickness of 1 μm.

With the GEANT4 simulations, an evaluation of detector response to background radiation and incident ions was performed. The simulated results obtained from the irradiation with photons indicate that the detector sensitivity was of $10^{-6}$, i.e. four orders of magnitude less than the previously used scintillator detectors. In addition, the cluster size for the interaction of the Compton electrons in the sensitive volume was found to be of the order of 10 μm.

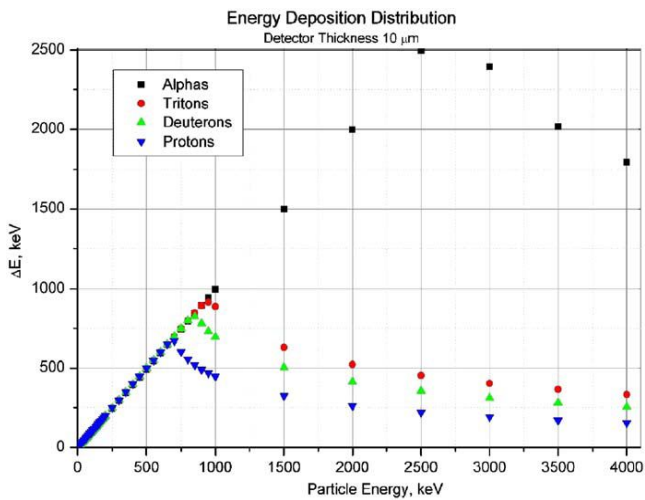

Fig. 2. Energy deposition for alpha particles, tritons, deuterons and protons in a wide energy range.

Similar simulations were performed for incident ions: alpha particles, deuterons, tritons and protons. In the figure below, the energy deposition of ions for a wide range of the incident energy is shown. It was found that the deposited energy has similar behavior for all the ions; it first shows linear increase with the incident particle energy, then a saturation point and, finally, with further increase of the incident particle energy smaller energy is deposited. As we can see in Fig. 2, the saturation point is reached at different energy levels for different types of particles, e.g. alpha particles reach the saturation at 2.5 MeV, then tritons at even lower with 950 keV, then the deuterons with 850 keV and the lowest ones were the protons with 700 keV.

### III. FABRICATION

A fabrication run was done in order to obtain the detector response (see Fig. 3). The main step in the realization of the U3DTHIN is to combine the fabrication technology of standard 3D detectors with thinning of planar devices. The process of drilling the holes to make the cylindrical contacts into the silicon substrate is done using an Inductively Coupled Plasma process by an Alcatel 601E dry etching machine.

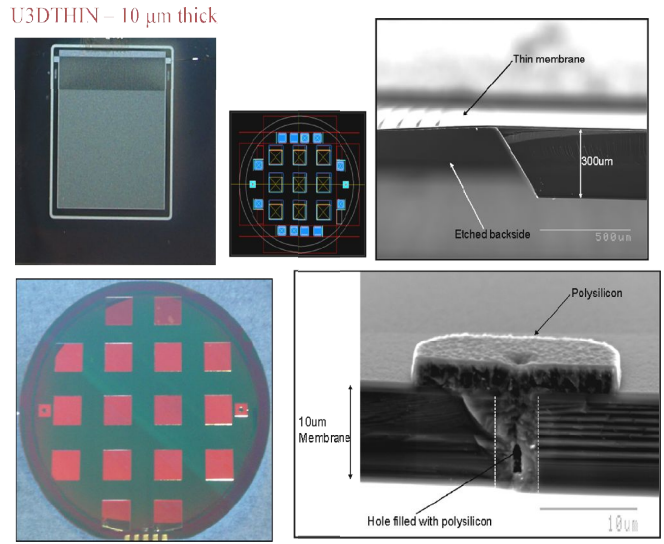

Fig. 3. Wafer with U3DTHIN detectors set of mask and SEM images

The process was optimized to stop the etching at the oxide interface of the Silicon on Insulator (SOI) wafer. In this run two types of holes were drilled, filled with polysilicon and then doped. Since these devices show to be very robust after filling all 3 μm holes, then it was proceed to deposited metal on the back surface to form electrical contact for the bias voltage and readout channels.

The final step in this fabrication process is thinning of the front surface, which will be the active detector area. The thinning is done using a TMAH solution, which stops etching at the oxide interface of the SOI wafer. This oxide is etched and then deposited with an atomic layer deposition (ALD).

In addition to that, adding a Boron conversion layer on top of the entrance window makes possible the detection of thermal neutrons.

### IV. IRRADIATION WITH AN ALPHA SOURCE

A front-end electronics with a charge sensitive pre-amplifier (See Fig. 4) was used and then the signals were linearly amplified before to be fed into a Multi channel analyzer.

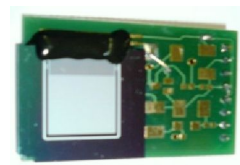

Fig. 4. U3DTHIN connected to the Pre-Amplifier.

On top of the detector an $^{241}$Am source was placed, in order to irradiate with alpha particles, the main idea was to try to get the first signals from the U3DTHIN (See Fig. 5). The distance between the source and the detector was 20,00 mm, which will means an average energy of the alpha particles of about 3,47 MeV. With this energy of the primary particles, we were quite confident that the detector was traversed and what we can obtain are signals that do not represent the full absorption.

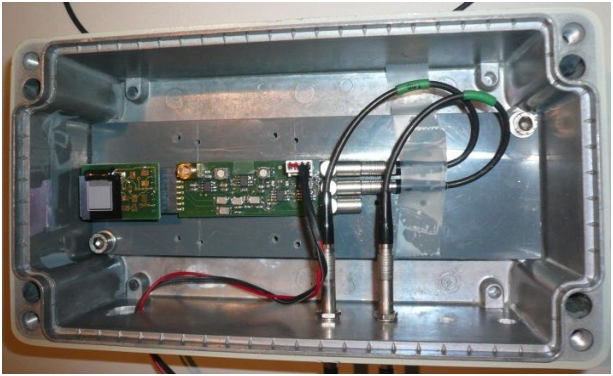

Fig. 5. Charge sensitive pre-Amplifier connected to a U3DTHIN and linear amplifier.

In the Fig. 6 the pulse high distribution of the signals obtained from the irradiation with an $^{241}$Am source is shown.

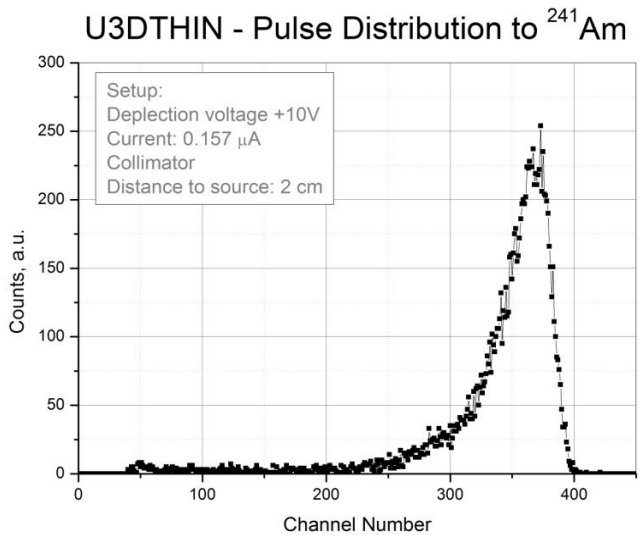

Fig. 6. Pulse high distribution of signals for an alpha source of the U3DTHIN.

As preliminary conclusion can be drawn and is that, the concept of the U3DTHIN has been tested and first signals have been obtained. A deeper study will be carried out and a calibration will be performed as well, this will help to establish the lower particle detection energy threshold.

ACKNOWLEDGMENT

This work is supported and financed by the Spanish Ministry of Education through the particle Physics National Program (ref. FPA2006-13238-C02-02) and co-financed with FEDER funds and through the GICSERV program "Access to ICTS integrated nano- and microelectronics clean room".